\documentclass[12pt, a4paper]{article}
\usepackage{amsmath,amssymb,amsfonts}
\usepackage{graphicx}
\usepackage{epstopdf}

\begin{document}

\title{Casimir energy and  force induced\\ by an impenetrable flux tube of finite radius}

\author{Volodymyr M. Gorkavenko
\thanks{E-mail: gorka@univ.kiev.ua}\\
\it \small Department of Physics, Taras Shevchenko National
University of Kyiv,\\ \it \small 64/13 Volodymyrska str., Kyiv
01601, Ukraine\\\phantom{11111111111}\\
Yurii A. Sitenko\thanks{E-mail: yusitenko@bitp.kiev.ua}, Olexander
B. Stepanov\thanks{E-mail:
\_\,pnd\_\,@ukr.net}\\
\it \small Bogolyubov Institute for Theoretical Physics,
\it \small National Academy of Sciences of Ukraine,\\
\it \small 14-b Metrologichna str., Kyiv 03680, Ukraine\\
\phantom{11111111111}}
\date{}

\maketitle

\begin{abstract}
A perfectly reflecting (Dirichlet) boundary condition at the edge of
an impenetrable magnetic-flux-carrying tube of nonzero transverse
size is imposed on the charged massive scalar matter field which is
quantized outside the tube. We show that the vacuum polarization
effects outside the tube give rise to a macroscopic force acting at
the increase of the tube radius (if the magnetic flux is held
steady). The Casimir energy and force are periodic in the value of
the magnetic flux, being independent of the coupling to the
space-time curvature scalar. We conclude that a topological defect
of the vortex type can polarize the vacuum of only those quantum
fields that have masses which are much less than a scale of the
spontaneous symmetry breaking.

\end{abstract}


\section{Introduction}

Polarization of the vacuum of quantum matter fields under the
influence of boundary conditions was studied intensively over more
than six decades since Casimir \cite{Cas} predicted a force between
grounded metal plates: the prediction was that the induced vacuum
energy in bounded spaces gave rise to a macroscopic force between
bounding surfaces, see reviews in Refs.~\cite{Mil} and \cite{Bor}.
The Casimir force between grounded metal plates has now been
measured quite accurately and agrees with the theoretical
predictions, see, e.g. Refs.~\cite{Lam} and \cite{Bre}, as well as
other publications cited in Refs.~\cite{Mil} and \cite{Bor}.

In the present paper we consider the vacuum energy which is induced
by boundary conditions in space that is not bounded but, instead, is
not simply connected, being an exterior to a straight infinitely
long tube. This setup is inspired by the famous Aharonov-Bohm effect
\cite{Aha}, and we are interested in polarization of the vacuum
which is due to imposing a boundary condition at the edge of the
tube carrying magnetic flux lines inside itself; this may be denoted
as the Casimir-–Aharonov-Bohm effect (see also \cite{Sit}).

The vacuum polarization effects which are due to imposing boundary
conditions of various types at the cylindrical surfaces were
extensively discussed in the literature, see \cite{Raad} --
\cite{Cavero}. In general, the Casimir effect in the presence of a
single smooth object (cylinder or sphere) is rather different from
that in the presence of two disjoint ones (e.g., plates): new
divergences appear, and to tame them one has to sum contributions of
quantized matter from both sides of the boundary surface, still this
does not help  to get rid completely of divergences,
see \cite{Bor} and references therein. In view of this, the
conventional prescription which is to subtract vacuum energy of
empty Minkowski space-time becomes insufficient for obtaining the
meaningful results. Some authors \cite{Wi1,Wi2} assert that there is no Casimir effect at all in this case.
 Our concern will not be in the case of an empty tube but, instead, in the case of a tube filled with the magnetic flux lines. We shall follow the author of ref.\cite{Schad} who  proposes to
define the Casimir energy for physical systems divided into classes:
the difference in vacuum energy of any two systems within the same
class should be finite, then the \textit{finite} Casimir energy has
the universal interpretation as a vacuum energy with respect to the
vacuum energy of a certain reference system which is common for the
whole class. We define a class of
physical systems corresponding to the charged matter field which is
quantized outside an impenetrable tube with the magnetic flux taking
different values; the case of zero flux can be chosen as the
reference system. As we shall show, the Casimir energy for this
class is unambiguous and finite.

A magnetic flux tube is formed inside a long current-carrying
solenoid or simply a magnetized whisker made of a ferromagnetic material, and its effect on the outside vacuum
can be studied in laboratory. Otherwise, a flux tube can be formed as a
topological defect of the vortex type, appearing after a phase
transition with spontaneous breakdown of the gauge symmetry
\cite{Abr,NO}: the condition of its appearance is that the first
homotopy group of the group space of the broken symmetry group be
nontrivial. The vortex is characterized by flux $2\pi\hbar c
e_H^{-1}$, where $e_H$ is the coupling constant of the Higgs scalar
field to the vortex-forming gauge field; the transverse size of the
vortex is of the order of correlation length $\hbar (m_H c)^{-1}$,
where $m_H$ is the mass of the Higgs scalar field. The issue of
vortices is widely discussed in condensed matter physics (e.g.
Abrikosov vortices in superconductors, see \cite{Hue}), as well as
in astrophysics and cosmology (e.g. cosmic strings, see
\cite{vilenkin,hind}). While considering the effect of the vortices
on the vacuum of the surrounding quantum matter, the following two
circumstances should be kept in mind: 1) the phase with broken
symmetry exists outside the vortex which is topological defect, and
the vacuum is to be defined only where the phase exists, hence the
quantum matter field does not penetrate inside the vortex, obeying a
boundary condition at its edge, 2) the coupling constant ($e$)  of
the quantum matter field to the vortex-forming gauge field differs,
in general, from $e_H$ (e.g. $e=e_H/2$ for normal excitations in
superconductors).

A simplifying assumption consists in a neglect of a transverse size
of the flux tube, i.e. in a use of an approximation of an infinitely
thin singular thread. Energy density and other components of the
energy-momentum tensor, which are induced in the background of a
singular magnetic thread, were studied in \cite{our2} (see also
\cite{Sit,Sit2}). The quantum matter field obeys the regularity
condition at the location of the thread, and the vacuum polarization
effects are periodic in the value of the magnetic flux with the
period equal to the London flux quantum ($2\pi\hbar c e^{-1}$); the
absolute value of the induced vacuum energy density is maximal at
half of the London flux quantum. A shortcoming of the approximation
of a singular thread is the power divergence of the induced vacuum
energy-momentum tensor in the vicinity of the thread, and, as a
consequence, neither Casimir energy (i.e. the induced vacuum energy
per unit length of the thread) nor Casimir force can be defined in
this approximation.

The transverse size of the magnetic flux tube was taken into account
in \cite{Dunne} -- \cite{Gra}, where it was shown that the induced
vacuum energy per unit length of the tube depends on the
configuration of the magnetic field inside the tube, being quadratic
in the flux for sufficiently smooth configurations. However, these
authors were concerned with the case when the region of the flux was
penetrable for the quantum matter field; therefore, their results
have no relation to the Casimir-Aharonov-Bohm effect. When the
quantum matter field is excluded from the region of the flux (that
is appropriate for the interpretation of the flux tube as a
topological defect), then the vacuum polarization effects become
independent of the details of the magnetic field configuration and
depend periodically on the whole flux \cite{newstring} -- \cite{newstring3}; meanwhile the contribution to both the Casimir energy and force which is due to the magnetic flux in the excluded region is well-defined. In the following quantum matter will be represented by the charged massive scalar field. As we shall see,
the vacuum energy which is induced outside the flux tube gives rise
to a macroscopic force acting at the increase of the tube radius, if
the magnetic flux is held steady. Although the induced vacuum energy
density depends on the coupling of the scalar field to the
space-time curvature scalar, the Casimir energy and force will be
shown to be independent of this coupling.

In the next section we define the renormalized induced vacuum energy
density in the background of an impenetrable flux tube and review
briefly the obtained earlier results as to its behavior in a plane,
i.e. when the spatial dimension along the tube is ignored. The
Casimir energy and force in a plane are considered in Section 3. The
longitudinal dimensions are added in Section 4 where we find the
Casimir energy and force in the most general case of a $(d-2)$-tube in
$d$-dimensional space. The obtained results are summarized and
discussed in Section 5.

\section{Vacuum energy density}

The temporal component of the energy-momentum tensor for the
quantized charged scalar field $\Psi(x)$ in flat space-time is given
by expression
\begin{equation}\label{t00}
T_{00}(x)=\frac12\left[\partial_0\Psi^\dag,\partial_0\Psi\right]_+-\frac14\left[\partial_0^2\Psi^\dag,\Psi\right]_+
-\frac14\left[\Psi^\dag,\partial_0^2\Psi\right]_+-\left(\xi-\frac14\right){\mbox{\boldmath
$\nabla$}}^2\left[\Psi^\dag,\Psi\right]_+,
\end{equation}
where ${\mbox{\boldmath $\nabla$}}$ is the covariant spatial
derivative involving both affine and bundle connections and the
field operator in the case of a static (ultrastatic) background
takes form
\begin{equation}\label{a11}
 \Psi(x^0,{\textbf{x}})=\sum\hspace{-1.4em}\int\limits_{\lambda}\frac1{\sqrt{2E_{\lambda}}}\left[e^{-{\rm i}E_{\lambda}x^0}\psi_{\lambda}({\bf x})\,a_{\lambda}+
  e^{{\rm i}E_{\lambda}x^0}\psi_{-\lambda}({\bf
  x})\,b^\dag_{\lambda}\right];
\end{equation}
units $\hbar=c=1$ are used, $a^\dag_\lambda$ and $a_\lambda$
($b^\dag_\lambda$ and $b_\lambda$) are the scalar particle
(antiparticle) creation and destruction operators satisfying
commutation relations; wave functions $\psi_\lambda(\textbf{x})$
form a complete set of solutions to the stationary Klein-Gordon
equation
\begin{equation}\label{a12}
 \left(-{\mbox{\boldmath $\nabla$}}^2  + m^2\right)  \psi_\lambda({\bf x})=E^2_\lambda\psi({\bf x}),
\end{equation}
 $m$ is the mass of the scalar
particle; $\lambda$ is the set of parameters (quantum numbers)
specifying the state; $E_\lambda=E_{-\lambda}>0$ is the energy of
the state; symbol
  $\sum\hspace{-1em}\int\limits_\lambda$ denotes summation over discrete and
  integration (with a certain measure) over continuous values of
  $\lambda$.

As is known for a long time \cite{Pen,Cher,Cal}, the energy-momentum
tensor depends on parameter $\xi$ which couples $\Psi$ to the scalar
curvature of space-time even in the case of the vanishing curvature,
see \eqref{t00}; conformal invariance is achieved in the limit of
vanishing mass $(m=0)$ at $\xi=(d-1)(4d)^{-1}$, where $d$ is the
spatial dimension. Consequently, the density of the induced vacuum
energy which is given formally by expression
\begin{equation}\label{a14}
   \varepsilon=\langle {\rm vac}|T_{00}(x)|{\rm vac} \rangle=\sum\hspace{-1.4em}\int\limits_{\lambda}E_\lambda\psi^*_\lambda(\textbf{x})\,\psi_\lambda(\textbf{x})-(\xi-1/4)
   {\mbox{\boldmath $\nabla$}}^2
  \sum\hspace{-1.4em}\int\limits_{\lambda}E^{-1}_\lambda\psi^*_\lambda(\textbf{x})\,\psi_\lambda(\textbf{x})
\end{equation}
depends on $\xi$ as well. This poses a question: whether physically
measurable effects (e.g. the Casimir force) can be dependent on
$\xi$?

In the present paper we are considering a static background in the
form of the cylindrically symmetric magnetic flux tube of finite
transverse size, hence the covariant derivative is $\mbox{\boldmath
$\nabla$}=\mbox{\boldmath $\partial$}-{\rm i}e {\bf V}$ with the
vector potential possessing only one nonvanishing component given by
\begin{equation}\label{3}
V^\varphi=\Phi/2\pi
\end{equation}
outside the tube; here $\Phi$ is the value of the magnetic flux and
$\varphi$ is the angle in  polar $(r,\varphi)$ coordinates on a
plane which is transverse to the tube.

The vacuum polarization  depends on the choice of a boundary
condition at the edge of the tube $(r=r_0)$. We impose, as in
\cite{newstring} -- \cite{newstring3}, the Dirichlet boundary
condition:
\begin{equation}\label{4}
\left.\psi_\lambda\right|_{r=r_0}=0,
\end{equation}
i.e. quantum matter is assumed to be perfectly reflected from the
thence impenetrable flux tube. Other possible choices  of a boundary
condition will be considered elsewhere.

The solution to \eqref{a12} outside the magnetic flux tube can be
obtained in terms of the cylindrical functions. The formal
expression \eqref{a14} for the vacuum energy density has to be
renormalized by subtracting the contribution corresponding to the
zero flux. The tube in 3-dimensional space can be obviously
generalized to the $(d-2)$-tube in $d$-dimensional space by adding
extra $d-3$ dimensions as longitudinal ones. Thus
 we obtain (for details see \cite{newstring,newstring2}):
\begin{multline}\label{c2}
\varepsilon_{ren}=(2\pi)^{1-d} \int d^{d-2}{\bf
k}_z\int\limits_0^\infty
  dk\,k\left(\sqrt{ {\bf k}_z^2+k^2+m^2}-\frac{\xi-1/4}{\sqrt{{\bf k}_z^2+k^2+m^2}}\triangle\right)\times\\ \times\left[S(kr,kr_0)-S(kr,kr_0)|_{\Phi=0}\right],
\end{multline}
 where
\begin{equation}\label{a29a}
 S(kr,kr_0)=\sum_{n\in\mathbb
 Z}\frac{\left[Y_{|n-e\Phi/2\pi|}(kr_0)J_{|n-e\Phi/2\pi|}(kr)-J_{|n-e\Phi/2\pi|}(kr_0)Y_{|n-e\Phi/2\pi|}(kr)\right]^2}{Y^2_{|n-e\Phi/2\pi|}(kr_0)+J^2_{|n-e\Phi/2\pi|}(kr_0)},
\end{equation}
$\mathbb{Z}$ is the set of integer numbers, $J_\mu(u)$ and
$Y_\mu(u)$ are the Bessel and the Neumann functions of order $\mu$,
the integration over the components of the ($d-2$)-dimensional
momentum ${\bf k}_z$ ranges from $-\infty$ to $\infty$, and
$\triangle=\partial^2_r+r^{-1}\partial_r\,$ is the
 radial part of the Laplacian operator on the plane which is
 orthogonal to the ($d-2$)-tube.

Owing to the infinite range of summation, the last expression is
periodic in flux $\Phi$ with a period equal to $2\pi e^{-1}$, i.e.
the London flux quantum (in units $c=\hbar=1$). Our further analysis
concerns the case of $\Phi=(2n+1)\pi e^{-1}$ when each of the
integrals in \eqref{c2} is the most distinct from zero. Introducing
function
\begin{equation}\label{21}
G(kr,kr_0)=S(kr,kr_0)|_{\Phi=\pi e^{-1}}-S(kr,kr_0)|_{\Phi=0},
\end{equation}
we rewrite \eqref{c2} in the case of $d=2$ in the dimensionless form
\begin{equation}\label{1}
r^3\varepsilon_{ren}=\alpha_+(mr_0,mr)-(\xi-1/4)r^3\triangle
\frac{\alpha_-(mr_0,mr)}{r},
\end{equation}
where
\begin{equation}\label{c3ab}
\alpha_\pm(mr_0,mr)=\frac{1}{2\pi}\int\limits_0^\infty
dz\,z\left[z^2+\left(\frac{mr_0}\lambda\right)^2\right]^{\pm1/2}
G(z,\lambda z),
\end{equation}
and  $\lambda=r_0/r$ ($\lambda\in[0,1]$).

Functions $\alpha_+$ and $\alpha_-$ were numerically calculated at a
set of different distances from the axis of the tube in
\cite{newstring2,newstring3} where it was shown that the results can
be approximated by the interpolation function in the form
\begin{equation}\label{m1}
\alpha_\pm(x_0,x)=\left[\pm e^{-2x}x^{1\mp1/2}\right]
\left[\left(\frac{x-x_0}{x}\right)^2\frac{P^\pm_3(x-x_0)}{x^3}\right]
\frac{Q^\pm_3(x^2)}{R^\pm_3(x^2)},\quad x>x_0,
\end{equation}
where $x=mr$, $x_0=mr_0$ and $P^\pm_n(y)$, $Q^\pm_n(y)$,
$R^\pm_n(y)$  are polynomials in $y$ of the $n$-th order with the
$x_0$-dependent coefficients. First factor in square bracket in
\eqref{m1} describes the large distance behavior  in the case of the
zero-radius tube (singular thread), second factor in  square bracket
is an asymptotics at small distances from the edge of the tube, and
the last quotient is the intermediate part. Since
the flux tube is impenetrable, the $\alpha_\pm$ functions vanish at
$x\leq x_0$.

For the $\alpha_+$ function  we estimate the relative error of the
obtained result as $0.1\%$. It should be noted that nearly 95 \% of
the integral value  is obtained by direct calculation and only
nearly 5\% is the contribution from the interpolation. The
integration in the case of the $\alpha_-$ function is performed more
quickly and with a higher accuracy, as compared to the case of the
$\alpha_+$ function, because the former tends to zero more rapidly
at large distances. In this case the contribution from the
interpolation can be estimated as $10^{-3}\%$ from the total value.

We define function \cite{newstring2}
\begin{equation}\label{m2}
\tilde
\alpha_-(x_0,x)=r^3\triangle\frac{\alpha_-(x_0,x)}{r}=\alpha_-(x_0,x)-x\frac{\partial
\alpha_-(x_0,x)}{\partial x}+x^2\frac{\partial^2
\alpha_-(x_0,x)}{\partial x^2}
\end{equation}
and construct the dimensionless vacuum energy density  at different
values of the coupling to the space-time curvature scalar ($\xi$) in
the form:
\begin{equation}\label{m3}
r^3\varepsilon_{ren}=\alpha_+(x_0,x)-(\xi-1/4)\tilde\alpha_-(x_0,x).
\end{equation}

The behavior of $\alpha_\pm$ and $\tilde\alpha_-$ and
$r^3\varepsilon_{ren}$ as functions of the distance from the axis of
the tube for different values of $r_0$ and $\xi$ was analyzed in
\cite{newstring,newstring2}. Of primary interest is the behavior at
the decrease of the tube radius. It seems plausible that this case
becomes more similar to the case of the tube of zero radius
(singular thread). However there are some peculiarities in the
behavior in the vicinity of the tube, and we discuss them following
\cite{newstring3}. Let us first recall the exact expressions
corresponding to the case of the singular magnetic thread (see
\cite{our2}):
\begin{align}\label{singa}
%&\alpha_+(0,x)=\frac{x^3}{3\pi^2}\left(\frac\pi2+\frac{K_0(2x)}{2x}-\left(1-\frac{1}{2x^2}\right)K_1(2x)-\right.\nonumber\\
%&\hspace{16.6em}\left.\vphantom{\frac12}-\pi
%x \left[K_0(2x)L_{-1}(2x)+K_1(2x)L_{0}(2x)\right]\right),\\
&\alpha_+(0,x)=\frac{x^3}{3\pi^2}\left\{\frac\pi2-2x
K_0(2x)-K_1(2x)+\frac{K_2(2x)}{2x}-\right.\nonumber\\
&\hspace{16.9em}\left.\vphantom{\frac12}-\pi x
\left[K_0(2x)L_{1}(2x)+K_1(2x)L_{0}(2x)\right]\right\},\\
&\alpha_-(0,x)=\frac{x}{\pi^2}\left\{\frac\pi2-2x
K_0(2x)-K_1(2x)-\pi
x \left[K_0(2x)L_{1}(2x)+K_1(2x)L_{0}(2x)\right]\right\},\\
&\tilde\alpha_-(0,x)=-\frac{x}{\pi^2}[2x K_0(2x)+K_1(2x)],
\end{align}
where $K_\nu(u)$ and $L_\nu(u)$ are the Macdonald and the modified
Struve functions of  order $\nu$. Consequently, in the vicinity of a
thread one gets
\begin{align}\label{singasmall1}
&\alpha_+(0,x)=\frac{1-3x^2}{12\pi^2},\quad x\ll 1\\
&\alpha_-(0,x)=-\frac{
1 - \pi x + (3  - 2 \gamma - 2 \ln x)x^2}{2\pi^2}+O(x^3),\quad x\ll 1,\label{singasmall2}\\
&\tilde\alpha_-(0,x)=-\frac1{2\pi^2}+\frac{1+2\gamma+2\ln
x}{2\pi^2}\,x^2+O(x^3),\quad x\ll 1,\label{singasmall3}
\end{align}
where $\gamma$ is the Euler constant. Using the latter relations, we
get asymptotics of the renormalized vacuum energy density at small
distances from the singular magnetic thread
\begin{equation}\label{g8}
r^3\varepsilon_{ren}^{sing}=\frac1{12\pi^2}-\frac{x^2}{4\pi^2}-\left(\xi-\frac14\right)\left(-\frac{1}{2\pi^2}+\frac{1
+ 2 \gamma + 2 \ln x }{2\pi^2}x^2\right)+O(x^3),\quad x\ll1.
\end{equation}

In contrast to \eqref{singasmall1} and \eqref{singasmall2}, the
$\alpha_\pm(x_0,x)$ functions in the case of nonzero radius are
vanishing quadratically in the vicinity of the tube, see
\cite{newstring2},
\begin{equation}\label{g8a}
\left.\alpha_\pm(x_0,x)\right|_{x\rightarrow x_0}\sim
O\left[(x-x_0)^2\right].
\end{equation}
To be more precise, we assume the asymptotics in the form, cf.
\eqref{m1},
\begin{equation}\label{alpm}
\alpha_\pm(x_0,x)=\pm\frac{(x-x_0)^2}{x^2}f_\pm(x_0,x),
\end{equation}
then one gets
\begin{multline}\label{dop7}
\tilde\alpha_-(x_0,x)=-(x-x_0)^2\frac{\partial^2}{\partial
x^2}f_-(x_0,x)+\left(1-6\frac{x_0}{x}+5\frac{x_0^2}{x^2}\right)x\frac{\partial}{\partial
x}f_-(x_0,x)-\\-\left(1-8\frac{x_0}{x}+9\frac{x_0^2}{x^2}\right)f_(x_0,x),
\end{multline}
with $\tilde\alpha_-(x_0,x_0)=-2f_-(x_0,x_0)$.

\begin{figure}[t]
\begin{center}
\includegraphics[width=150mm]{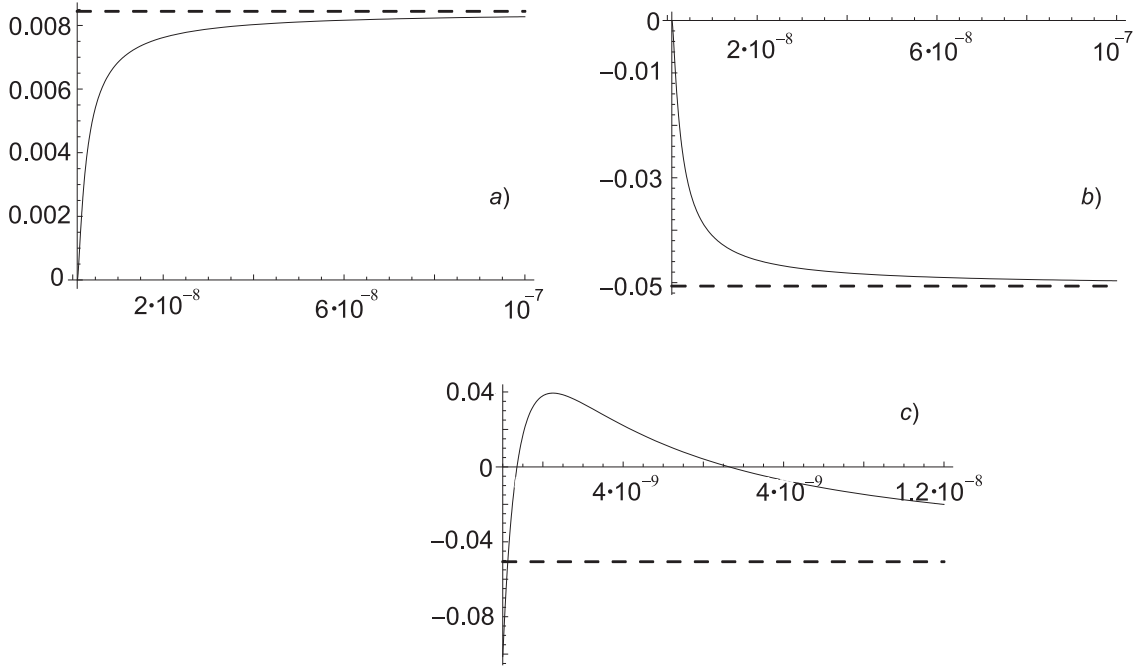}
\end{center}
\caption{The behavior of the constituents of the dimensionless
vacuum energy density at small distances from the tube: \textit{a})
$\alpha_+$, \textit{b}) $\alpha_-$, \textit{c}) $\tilde \alpha_-$
for the case of $x_0=10^{-9}$ (solid line). The behavior of the
corresponding functions for the case of a singular magnetic thread
is presented by a dashed line. Variable $x$ $(x>x_0)$ is along the
abscissa axis.\label{fig5}}
\end{figure}

The $f_\pm(x_0,x)$ functions are adjusted as
\begin{align}\label{001}
& f_+(0,x)=\frac{
1-3x^2}{12\pi^2},\quad x\ll1,\\
& f_-(0,x)=\frac{ 1 - \pi x + (3  - 2 \gamma - 2 \ln
x)x^2}{2\pi^2},\quad x\ll1\label{002};
\end{align}
consequently, one gets
\begin{multline}\label{003}
\left.\vphantom{\frac12}\tilde\alpha_-(x_0,x)\right|_{\!\!\!\scriptsize\begin{array}{l}
x_0\rightarrow 0\\ x\rightarrow x_0\end{array}}
=-\frac1{2\pi^2}+\frac{1+2\gamma+2\ln x}{2\pi^2}\,x^2+\frac{4-\pi
x}{\pi^2x}\,x_0+\\+\frac{-9 + 4 \pi x - 7 x^2 + 2 \gamma x^2 + 2 x^2
\ln x}{2\pi^2 x^2}\,x_0^2.
\end{multline}

The asymptotical behavior of the $\alpha_\pm$ and $\tilde\alpha_-$
functions with the use of \eqref{alpm} -- \eqref{003}  is presented
on Fig.1 for the case of a sufficiently small value of $x_0$.  It
should be noted that the $f_\pm(x_0,x)$ functions depend strongly on $x_0$.

\section{Total vacuum energy and the Casimir force in a plane}

The total vacuum energy which is induced in a plane outside the
magnetic flux region is
\begin{equation}\label{g3a}
E_2=2\pi m \left[\int\limits_{x_0}^\infty
\frac{\alpha_+(x_0,x)}{x^2}\,dx-\left(\xi-\frac14\right)\int\limits_{x_0}^\infty
\frac{\tilde\alpha_-(x_0,x)}{x^2}\,dx\right].
\end{equation}
In view of the relation
\begin{equation}\label{m5}
\int\limits_{x_0}^\infty \frac{\tilde\alpha_-(x_0,x)}{x^2}\,
dx=-x\left.\frac{\partial}{\partial
x}\left(\frac{\alpha_-(x_0,x)}{x}\right)\right|_{x=x_0}
\end{equation}
which follows from \eqref{m2}, and relations \eqref{alpm} and
\eqref{002}, we conclude that the total vacuum energy is independent
of the coupling to the space-time curvature scalar ($\xi$):
\begin{equation}\label{dop1}
E_2= m \mathcal{D}(mr_0),
\end{equation}
where
\begin{equation}\label{dop1a}
 \mathcal{D}(x_0)=2\pi \int\limits_{x_0}^\infty
\frac{\alpha_+(x_0,x)}{x^2}\,dx.
\end{equation}
This is in contrast to the case of the singular magnetic thread,
when the total induced vacuum energy is divergent and
$\xi$-dependent (see \cite{our2}):
\begin{equation}\label{dop2}
E^{sing}_2\equiv\int\limits_{0}^{2\pi}d\varphi\int\limits_{0}^\infty
\varepsilon_{ren}^{sing} \,r dr\sim
4m\left(\xi-\frac1{12}\right)\int\limits_0\frac{dx}{x^2}.
\end{equation}
It is curious that the vacuum energy in this case is finite at
$\xi=1/12$, being equal to
\begin{multline}\label{dop3a}
\left.E^{sing}_2\right|_{\xi=1/12}=\frac{2m}{3\pi}\int\limits_0^\infty\left\{\frac\pi2-\left(2x+\frac1{2x}\right)K_0(2x)-K_1(2x)-\right.\\
-\left.\vphantom{\frac12}\pi
x\left[K_0(2x)L_1(2x)+K_1(2x)L_0(2x)\right]\right\}x\,dx=-0.01989\times
2\pi m,
\end{multline}
and taking the negative value.

Although vacuum energy $E_2$ \eqref{dop1} is finite, its value grows
infinitely as $x_0$ tends to zero (see \eqref{alpm} and
\eqref{001}):
\begin{equation}\label{dop3}
\left.E_2\right|_{x_0\rightarrow0}=m\left[\frac{1}{18\pi
x_0}-\frac{x_0}{\pi}\ln x_0+O(x_0^3)\right],
\end{equation}
which is in accordance with the divergence of the vacuum energy in
the case of the singular magnetic thread. To be more precise,
relation \eqref{m5} fails to yield zero in the case $x_0=0$, and,
therefore, the divergence of the vacuum energy in the latter case
becomes $\xi$-dependent.

\begin{figure}[t]
\begin{center}
\includegraphics[width=160mm]{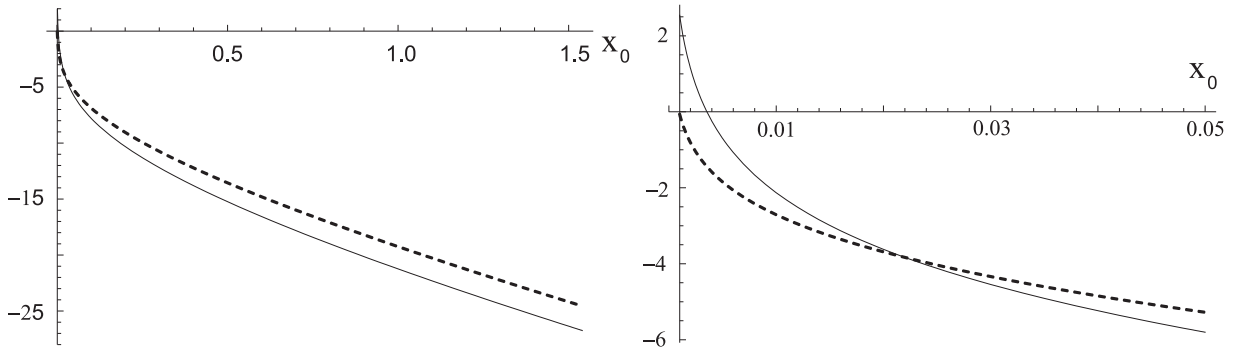}
\end{center}
\caption{The logarithm of the induced vacuum energy in the plane as
a function of the tube radius starting from $x_0=10^{-3}$:
$\ln\frac{E_2}{ m}$ is given by a dashed curve and
$\ln\frac{E_3}{m^2}$ (see Section 4) is given by a solid curve.
\label{fig6}}
\end{figure}

We present the values of vacuum energy $E_2$ \eqref{dop1} for
several values of the tube radius in the second  row of the Table.
\begin{center}
\begin{tabular}{|c|c|c|c|c|c|c|}
\hline  $m r_0$ & 3/2 & 1 & $1/2$ & $10^{-1}$ & $10^{-2}$ & $10^{-3}$\\
\hline $E_2/m$ & $3.15\cdot 10^{-11}$ & $4.363\cdot10^{-9}$
& $1.299\cdot10^{-6}$ & $1.038\cdot 10^{-3}$ &
$0.0666$ & $0.933$\\
\hline $E_3/ m^2$ & $3.577\cdot 10^{-12}$ &
$5.942\cdot10^{-10}$ & $2.411\cdot10^{-7}$ & $4.162\cdot10^{-4}$ &
$0.119$ & $12.704$\\
\hline
\end{tabular}
\end{center}

{\small Table 1. Values of the dimensionless vacuum energy at several values of $mr_0$.}

\phantom{gxdhh}

 These results are also given
on Fig.2 in a logarithmic scale, where the dots corresponding to the
data in the Table are joined with the help of an interpolation
function $\eta(x_0)=\ln \frac{E_2}{m}$,
%\begin{equation}\label{dop4}
%\eta(x_0)=\ln \frac{E_2}{2\pi m},
%\end{equation}
which, for the range $x_0>10^{-3}$, can be taken in the form
\begin{equation}\label{g2}
\eta(x_0)=-a-x_0^{b}P_n(x_0)-\left(c+x_0^{d}Q_n(x_0)\right)\ln x_0.
\end{equation}
where $a,b,c,d$  are the positive constants and $P_n(x_0),Q_n(x_0)$
are polynomials in $x_0$ of the $n-$th order.

To change the radius of the  flux tube one has to apply a work that
is equal to the change of the total vacuum energy which is induced
outside the tube. In the case of the infinitely small change of the
radius one has
\begin{equation}\label{g4}
\delta E_2=2\pi P_2\, r_0 \delta r_0 ,
\end{equation}
where $P_2$ can be interpreted as the vacuum pressure which acts
from the outside to the inside of the tube
\begin{equation}\label{g5}
P_2=\frac{1}{2\pi r_0}\frac{{\rm d} E_2}{{\rm d} r_0}=
\frac{m^3}{2\pi x_0}\mathcal{D}'(x_0),
\end{equation}
$\mathcal{D}'(x_0)=\frac{{\rm d}}{{\rm d} x_0}\mathcal{D}(x_0)$, and
the value of the magnetic flux inside the tube is assumed to remain
unchanged.

 This results in the
Casimir force acting from the inside to the outside of the tube
\begin{equation}\label{g6}
F_2=-2\pi r_0 P_2=-m^2\mathcal{D}'(x_0).
\end{equation}
As the tube radius tends to zero, the Casimir force grows
infinitely:
\begin{equation}\label{dddop1}
F_2=m^2\left(\frac{1}{18\pi x_0^2}-\frac{1}{\pi}(\ln
x_0+1)+O(x_0^2)\right).
\end{equation}
 The behavior of the Casimir force is presented on Fig.3.

\begin{figure}[t]
\begin{center}
\includegraphics[width=160mm]{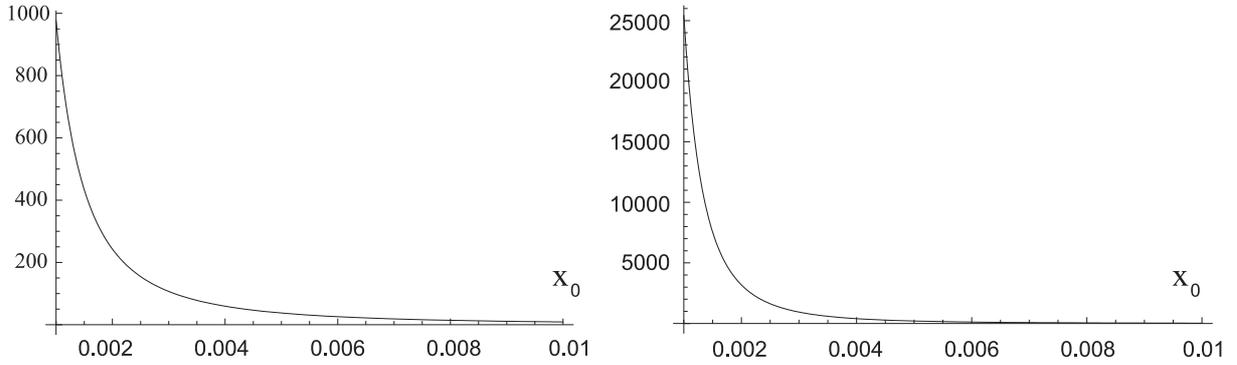}
\end{center}
\caption{The Casimir force as a function of the tube radius in the
range $10^{-3}<x_0<10^{-2}$: $F_2/m^2$ is on the left
and $F_3/m^3$ (see Section 4) is on the right.
\label{fig7}}
\end{figure}

As one can see, the Casimir force tends to increase the radius of
the tube and to minimize the induced vacuum energy of the quantized
scalar field. As to the energy stored inside the tube, it is the
purely classical energy of the magnetic field. Its behavior at the
increase of the tube radius as the magnetic flux is held constant
can be different depending on the details of the magnetic field
configuration. Mild assumptions as to the smoothness of the
configuration yield that the energy is either constant or decreasing
at most as $\sim r_0^{-2}$.

\section{Generalization to higher than two dimensions}

In $d$-dimensional space, we define the vacuum energy which is
induced outside a $(d-2)$-tube in a plane orthogonal to it:
\begin{equation}\label{4s1}
E_d=\int\limits_{0}^{2\pi}\,d\varphi\, \int\limits_{r_0}^\infty
dr\,r\varepsilon_{ren},
\end{equation}
where $\varepsilon_{ren}$ is given by \eqref{c2}. Generalizing
relation \eqref{m5} we obtain relation
\begin{multline}\label{4s2}
\int\limits_{r_0}^\infty dr\,r\Delta \int d^{d-2}{\bf
k}_z\int\limits_0^\infty \frac{dk\, k}{\sqrt{{\bf
k}_z^2+k^2+m^2}}\left[S(kr,kr_0)-S(kr,kr_0)|_{\Phi=0}\right]=\\=\left.
-\left\{r\partial_r \int d^{d-2}{\bf k}_z\int\limits_0^\infty
\frac{dk\, k}{\sqrt{{\bf
k}_z^2+k^2+m^2}}\left[S(kr,kr_0)-S(kr,kr_0)|_{\Phi=0}\right]\right\}\right|_{r=r_0}.
\end{multline}
The right-hand side of \eqref{4s2} is obviously vanishing due to
relation
\begin{equation}\label{4s3}
[r\partial_r S(kr,kr_0)]|_{r=r_0}=0,
\end{equation}
stemming from the definition of $S(kr,kr_0)$, see \eqref{a29a}.
Consequently, the Casimir energy, i.e. the induced vacuum energy per
unit length of the $(d-2)$-tube, is independent of the coupling to
the space-time curvature scalar
\begin{equation}\label{4s4}
E_d=(2\pi)^{2-d}\int\limits_{r_0}^\infty dr \,r \int d^{d-2}{\bf
k}_z\int\limits_0^\infty
  dk\,k\sqrt{{\bf k}_z^2+k^2+m^2}\left[S(kr,kr_0)-S(kr,kr_0)|_{\Phi=0}\right].
\end{equation}
Changing the order of integration over $r$ and ${\bf k}_z$, we
relate $E_d$ to the total induced vacuum energy in the $d=2$ case,
$E_2$ \eqref{dop1}:
\begin{equation}\label{4s5}
E_d=m^{d-1}\frac{(4\pi)^{1-d/2}}{\Gamma(d/2)}\int\limits_0^\infty
du\,\sqrt{1+u^{2/(d-2)}}\,
\mathcal{D}\left(x_0\sqrt{1+u^{2/(d-2)}}\right),
\end{equation}
where $u=(|{\bf k}_z|r_0)^{d-2}$,  $\Gamma(y)$ is the Euler gamma function, and
\begin{equation}\label{dddop2}
\mathcal{D}(y)=\int\limits_y^\infty\frac{dx}{x^2}\int\limits_0^\infty
dz\, z
\sqrt{z^2+x^2}\left[S\left(z,z\frac{y}{x}\right)-\left.S\left(z,z\frac{y}{x}\right)\right|_{\Phi=0}\right],
\end{equation}
is generalizing \eqref{dop1a} to arbitrary values of the flux.

Similarly to the $d=2$ case we define the Casimir force acting from
the inside to the outside of the $(d-2)$-tube along the radial
direction
\begin{equation}\label{4s6}
F_d=-\frac{{\rm d} E_d}{{\rm d} r_0},
\end{equation}
and relate it to the Casimir force in the $d=2$ case, $F_2$
\eqref{g6}:
\begin{equation}\label{4s7}
F_d=-m^d\frac{(4\pi)^{1-d/2}}{\Gamma(d/2)}\int\limits_0^\infty
du\left(1+u^{2/(d-2)}\right)\,\mathcal{D}'\left(x_0\sqrt{1+u^{2/(d-2)}}\right).
\end{equation}
It should be emphasized that relations \eqref{4s5} and \eqref{4s7}
are valid for arbitrary values of the flux. The finiteness of
integrals in \eqref{4s5} and \eqref{4s7} is due to the sufficiently
strong decrease of $\mathcal{D}(x_0)$ and $\mathcal{D}'(x_0)$ at
$x_0\gg1$, which was demonstrated for $\Phi=(2n+1)\pi e^{-1}$ in the
previous section.

Changing the integration variable in \eqref{4s5} and \eqref{4s7}, we
get
\begin{equation}\label{dddop3a}
E_d=\frac2{r_0^{d-1}}\frac{(4\pi)^{1-d/2}}{\Gamma\left(\frac{d-2}2\right)}\int\limits_{x_0}^\infty
dv\, v^2 \left(v^2-x_0^2\right)^{\frac{d-4}2}\,\mathcal{D}(v),
\end{equation}
\begin{equation}\label{dddop4}
F_d=-\frac2{r_0^d}\frac{(4\pi)^{1-d/2}}{\Gamma\left(\frac{d-2}2\right)}\int\limits_{x_0}^\infty
dv\, v^3 \left(v^2-x_0^2\right)^{\frac{d-4}2}\,\mathcal{D}'(v),
\end{equation}
where the latter in the case of $d>3$,  after integration by parts,
takes form
\begin{equation}\label{dddop5}
F_d=\frac2{r_0^d}\frac{(4\pi)^{1-d/2}}{\Gamma\left(\frac{d-2}2\right)}\int\limits_{x_0}^\infty
dv\, v^2
\left(v^2-x_0^2\right)^{\frac{d-6}2}[(d-1)v^2-3x_0^2]\,\mathcal{D}(v).
\end{equation}

At $x_0\ll 1$ we obtain
\begin{equation}\label{dddop6}
F_d=\frac{d-1}{r_0}E_d=\frac{C_\Phi(d)}{r_0^d}, \quad r_0\ll m^{-1},
\end{equation}
where
\begin{equation}\label{dddop7}
C_\Phi(d)=\frac{2(4\pi)^{1-d/2}}{\Gamma\left(\frac{d-2}2\right)}(d-1)\int\limits_0^\infty
dv\, v^{d-2}\,\mathcal{D}(v),\quad d>2,
\end{equation}
is monotonically decreasing with the increase of $d$. The numerical
estimate of $C(d)$ at $\Phi=(2n+1)\pi e^{-1}$ in the range $3\leq
d\leq 10$ yields that it can be well approximated as
\begin{equation}\label{dddop8}
C_\Phi(d)= (d-1)d^{10.025}\,{\rm
exp}\!\!\left(\frac{44.76}{d}-3d-28.097\right),
\end{equation}
where a decisive factor is $e^{-3d}$. In the following we shall use
a rough, but quite suitable for a further analysis, approximation
\begin{equation}\label{dddop8a}
C_\Phi(d)\approx (d-1)e^{-2.7d-4}
\end{equation}
that is valid for $d\leq10$.
In view of \eqref{dddop8a} and relation
\begin{equation}\label{dddop8b}
\frac1{d-2}\int\limits_{x_0}^\infty dv\, v
(v^2-x_0^2)^{\frac{d-2}2}\,\,[2\mathcal{D}'(v)+v\mathcal{D}''(v)]\leq
(d-1)\int\limits_0^\infty dv\, v^{d-2}\mathcal{D}(v),
\end{equation}
where the equality sign corresponds to sufficiently small values of
$x_0$, we find that the dimensionless force, $m^{-d}F_d$, as a
function of $d$ at $\Phi=(2n+1)\pi e^{-1}$ can be approximated as
\begin{equation}\label{dddop9}
m^{-d}F_d\approx \,(d-1)e^{-4}e^{-d(\ln x_0+2.7)}.
\end{equation}
Thus the dimensionless force increases with $d$ at $x_0\lesssim e^{-2.7}$ and
decreases with $d$ at $x_0\gtrsim e^{-2.7}$.

In the $d=3$ case we get
\begin{equation}\label{4s8}
E_3=\frac{m^2}{\pi} \int\limits_{0}^\infty du\,
\sqrt{1+u^2}\,\mathcal{D}\left(x_0\sqrt{1+u^2}\right)=-\frac{1}{\pi r_0^2}
\int\limits_{x_0}^\infty dv\,
\sqrt{v^2-x_0^2}\,[\mathcal{D}(v)+v\mathcal{D}'(v)]
\end{equation}
and
\begin{equation}\label{4s7a}
F_3=-\frac{m^3}{\pi}\int\limits_0^\infty
du\left(1+u^{2}\right)\,\mathcal{D}'\left(x_0\sqrt{1+u^{2}}\right)=\frac{1}{\pi r_0^3}\int\limits_{x_0}^\infty
dv\,v\sqrt{v^2-x_0^2}\, [2\mathcal{D}'(v)+v\mathcal{D}''(v) ].
\end{equation}
We present the values of Casimir energy $E_3$ \eqref{4s8} at
$\Phi=(2n+1)\pi e^{-1}$ $(n\in \mathbb{Z})$ for several values of
the tube radius in the third row of the Table. These results are
also given on Fig.2 in a logarithmic scale, where the dots
corresponding to the data in the Table are joined with help of an
interpolation function similarly to that as in the previous section;
the comparison is made with the $d=2$ case. Casimir force $F_3$
\eqref{4s7a} is presented on the right of Fig.3 and compared with
Casimir force $F_2$ \eqref{g6}; the former attains a considerable
value of $2.54\cdot 10^4\cdot  m^3$ at $r_0=10^{-3}m^{-1}$.

At $x_0\ll 1$, restoring constants $\hbar$ and $c$, we obtain
\begin{equation}\label{ddop9}
F_3=\frac{2}{r_0}E_3=\frac{\hbar c}{r_0^3}\,C_\Phi(3), \quad r_0\ll
m^{-1},
\end{equation}
where, see \eqref{dddop7},
\begin{equation}\label{dddop10}
C_\Phi(3)=\frac{2}{\pi}\int\limits_0^\infty dv\,v\mathcal{D}(v).
\end{equation}
Let us compare Casimir force $F_3$ \eqref{ddop9} with the force
caused by the classical magnetic field inside the tube. Assuming the
uniformity of the magnetic field filling completely the tube,
$B=\Phi/(\pi r_0^2)$, one obtains an expression for the classical
energy per unit length of the tube, $E^{(class)}=\Phi^2/(2\pi
r_0^2)$, which can be rewritten in terms of London flux quantum
$\Phi_0=2\pi \hbar c e^{-1}$ and fine structure constant
$\alpha=e^2(4\pi \hbar c)^{-1}$:
\begin{equation}\label{dddop11}
E^{(class)}=\frac{\hbar
c}{r_0^2}\frac{\Phi^2}{\Phi_0^2}\frac1{2\alpha}.
\end{equation}
The classical force which tends to decrease energy \eqref{dddop11}
by increasing the tube radius under the steady magnetic flux filling
completely the tube is
\begin{equation}\label{dddop12}
F^{(class)}=-\frac{{\rm d}}{{\rm d}r_0}E^{(class)}=\frac{\hbar
c}{r_0^3}\frac{\Phi^2}{\Phi_0^2}\frac1{\alpha}.
\end{equation}
Comparing this with Casimir force $F_3$ \eqref{ddop9}, we recall
that $C_\Phi(3)$ is a periodic function of the magnetic flux,
vanishing at $\Phi=n\Phi_0$. Even the maximal value of $C_\Phi(3)$
which is achieved at $\Phi=(n+1/2)\Phi_0$ and is equal to
$2.545\cdot10^{-5}$, see \eqref{dddop8a}, is more than million
times smaller than the value of the corresponding factor,
$\Phi^2/(\Phi_0^2\alpha)$, in \eqref{dddop12}: taking
$\Phi=\Phi_0/2$ one obtains value $(4\alpha)^{-1}\approx34.2$ for
this factor.

However, as it was already noted, the classical force acting from
the inside of the tube depends strongly on the detailed form of the
magnetic field configuration: it decreases if the magnetic field is
decreasing in the vicinity of the tube edge. For instance, in the
case of the magnetic field concentrated wholly inside a tube of
smaller radius, the classical force acting to extend the tube of
larger radius disappears at all, and only the Casimir force from the
outside vacuum is left in this capacity.

\section{Summary}
In the present paper we consider the vacuum polarization effects
which are induced in charged scalar matter by a magnetic flux enclosed in an impenetrable finite-radius tube; a perfectly reflecting (Dirichlet) boundary condition is imposed at the edge of the tube.
The previous analysis of the induced vacuum energy
density in the $d=2$ case \cite{newstring,newstring2} was extended
down to the values of the tube radius as small as $r_0=10^{-3}\hbar
(mc)^{-1}$ in \cite{newstring3}, where it was shown that contrary to
the case of a singular magnetic thread ($r_0=0$), the vacuum energy
density is finite everywhere, but its behavior is very similar to
that in the $r_0=0$ case, excepting the behavior in the vicinity of
the tube, where peculiar oscillations appear. The case of
$r_0<10^{-3}\hbar (mc)^{-1}$ is analyzed indirectly by combining the
numerical and analytical estimates, and the difference between the
$r_0=0$ and $r_0=10^{-9}\hbar (mc)^{-1}$ cases is illustrated by
Fig.1.

These two circumstances (the finiteness and at the same time the
similarity to the case of a singular thread) which are proven in the
$d=2$ case have far-reaching consequences that allow us to determine
the \textit{finite} Casimir energy in the case of space of arbitrary
dimension, as long as the tube radius is taken into account. We find
that the Casimir energy, i.e. the vacuum energy per unit length of
the $(d-2)$-tube, is positive and independent of the coupling to the
space-time curvature scalar $(\xi)$, notwithstanding the
$\xi$-dependence of the vacuum energy density and its lack of
positivity. The functional dependence of the Casimir energy on the
tube radius for the magnetic flux equal to half of the London flux
quantum is numerically estimated for the range
$10^{-3}\hbar(mc)^{-1}<r_0<1.5\hbar(mc)^{-1}$, and the results for
the $d=2$ and $d=3$ cases are presented in the Table and on Fig.2.
The Casimir energy is negligible for $r_0\sim \hbar(mc)^{-1}$, being
of order $10^{-10}\cdot m^{d-1}c^d\hbar^{2-d}$ for $d=2,3$ and
even less for larger $d$, but it increases considerably with the
decrease of the tube radius.

The Casimir energy gives rise to the Casimir force which is directed
from the inside to the outside of the tube along its normal. The
force is $\xi$-independent as well as the Casimir energy. The force
acts at the increase of the tube radius and the decrease of the
Casimir energy, if the  magnetic flux is held steady. The behavior
of the force as  a function of the tube radius in the $d=2$ and
$d=3$ cases for the magnetic flux equal to half of the London flux
quantum is illustrated by Fig.3. The force takes considerable values
at small values of the tube radius and actually disappears
otherwise:  in the $d=3$ case it is, e.g., $2.54\cdot 10^{4}\cdot
 m^3c^4\hbar^{-2}$ at $r_0=10^{-3}\hbar(mc)^{-1}$ and
$10^{-2}\cdot  m^3c^4\hbar^{-2}$ at $r_0=10^{-1}\hbar(mc)^{-1}$.

 It should be noted that we consider the case of the Casimir force
caused by a magnetic flux enclosed by a boundary where the Dirichlet boundary condition is imposed. The force is periodic in the flux value with a period
equal to the London flux quantum, attaining its maximal value at
$\Phi=(n+1/2)\Phi_0$ and vanishing at $\Phi= n\Phi_0$ ($n\in \mathbb
Z$). A general conclusion which is valid for arbitrary spatial
dimension $d\geq2$ is that the Casimir energy and force at $r_0\ll
\hbar(mc)^{-1}$, when they take considerable values, are actually
the same as they are in the case of the massless scalar field, see
\eqref{dop3}, \eqref{dddop1} and \eqref{dddop6}; the massive case
becomes formally distinct from the massless one at larger values of
the tube radius, when the Casimir energy and force take negligible
values. The Casimir force and energy increase with $d$ at smaller
$r_0$, when they are considerable, while decrease with $d$ at larger
$r_0$, when they are negligible, see \eqref{dddop9}; even the
comparison of numerical calculations for the $d=2$ and $d=3$ cases
reveals this fact, see the Table and Fig.2.

Whereas in the case of parallel plates the pure action of the
Casimir force to minimize the Casimir energy leads to a collapse,
the pure action of the Casimir force to minimize the Casimir energy
in the case of a flux tube leads not to a collapse but to an
expansion of the tube in the transverse direction. Note that the
classical energy of the constant magnetic flux inside the tube is
most likely to be constant or decreasing maximally as $r_0^{-2}$
with the expansion of the tube radius, see \eqref{dddop11}. Thus the
Casimir force tends to smear both quantum and classical effects of
the flux tube. The vacuum polarization is quite negligible at
$mcr_0>\hbar$, whereas it becomes noticeable at $mcr_0\ll \hbar$. If
the flux tube is interpreted as a topological defect of the vortex
type, then the vacuum polarization in its background is absent when
the mass of the Higgs field ($m_H\sim \hbar (r_0c)^{-1}$) does not
exceed the mass of the quantum matter field, $m_H\lesssim m$. Vacuum
polarization is essential for the quantum matter field with the mass
which is much less than the Higgs mass, $m\ll m_H$; since
$\Phi=2\pi\hbar c e_H^{-1}$ for the topological defect  case, the
effect is maximal when the coupling of the Higgs field to the gauge
field is twice the coupling of the quantum matter field to the gauge
field, $e_H=2e$ (e.g. the Higgs field describing the Cooper pair in
a superconductor). In particular, we can arrive at a conclusion that
a cosmic string which has been formed at the grand unification scale
polarizes the vacuum of the present-day quantum matter, but it has
no effect on the vacuum of matter fields with masses which are
comparable to the scale of grand unification.

\section*{Acknowledgments}

The work was partially supported by special program "Microscopic and
phenomenological models of fundamental physical processes in micro-
and macroworld" of the Department of Physics and Astronomy of the
National Academy of Sciences of Ukraine and by the ICTP ---
SEENET-MTP grant PRJ-09 ""Strings and Cosmology". V.M.G. and Yu.A.S. acknowledge the support from the State Agency for Science, Innovations and Informatization of Ukraine
under the SFFR-BRRFR grant F54.1/019.

\end{document}